\documentclass[twocolumn,floatfix,showpacs,superscriptaddress,prb]{revtex4}
\usepackage{graphicx}
\usepackage{amsmath}
\usepackage{bm}
\newcommand{\s}{\sum\limits}

\newcommand{\il}{\int\limits}
\newcommand{\be}{\begin{equation}}
\newcommand{\e}{\end{equation}}
\newcommand{\beml}{\begin{subequations}}
\newcommand{\eml}{\end{subequations}}
\newcommand{\beq}{\begin{eqnarray}}
\newcommand{\eq}{\end{eqnarray}}
\newcommand{\ba}{\begin{array}}
\newcommand{\ea}{\end{array}}
\newcommand{\lt}{\left}
\newcommand{\rt}{\right}
\newcommand{\n}{\nonumber}

\newcommand{\ep}{\varepsilon}

\begin{document}

\title{Excitation gap of a graphene channel with superconducting boundaries}
\author{M. Titov}
\affiliation{Department of Physics, Konstanz University, D--78457 Konstanz, Germany}
\author{A. Ossipov}
\affiliation{Instituut-Lorentz, Universiteit Leiden, P.O. Box 9506, 2300 RA Leiden, The Netherlands}
\author{C. W. J. Beenakker}
\affiliation{Instituut-Lorentz, Universiteit Leiden, P.O. Box 9506, 2300 RA Leiden, The Netherlands}
\date{September 2006}
\begin{abstract}
We calculate the density of states of electron-hole excitations in a
superconductor--normal-metal--superconductor (SNS) junction in
graphene, in the {\em long-junction\/} regime that the superconducting
gap $\Delta_{0}$ is much larger than the Thouless energy $E_{T}=\hbar
v/d$ (with $v$ the carrier velocity in graphene and $d$ the separation
of the NS boundaries). If the normal region is undoped, the excitation
spectrum consists of neutral modes that propagate along the boundaries
--- transporting energy but no charge. These ``Andreev modes'' are a
coherent superposition of electron states from the conduction band and
hole states from the valence band, coupled by specular Andreev
reflection at the superconductor. The lowest Andreev mode has an
excitation gap of $E_{0}=\frac{1}{2}(\pi-|\phi|)E_{T}$, with
$\phi\in(-\pi,\pi)$ the superconducting phase difference. At high
doping (Fermi energy $\mu\gg E_{T}$) the excitation gap vanishes
$\propto E_0 (E_{T}/\mu)^{2}$, and the usual gapless density of states of
Andreev levels is recovered. We use our results to calculate the
$\phi$-dependence of the thermal conductance of the graphene channel.
\end{abstract}
\maketitle

\section{Introduction}
\label{intro}

The two-dimensional layer of carbon atoms known as graphene is a
gapless semiconductor. A gap between conduction and valence bands
opens up if the layer is confined to a narrow channel.\cite{Sai98} For
a channel of width $d$ the band gap $2E_{0}$ is set by the (ballistic)
Thouless energy $E_{T}=\hbar v/d$, with $v$ the (energy independent)
velocity of electron and hole excitations in graphene. The size of the
gap depends on the crystallography of the channel edges. In
particular, for edges in the armchair configuration one
has\cite{Per06,Bre06}
\begin{equation}
E_{0}=\alpha E_{T}, \label{E0armchair}
\end{equation}
with $\alpha=0$ if the channel is a multiple of three unit cells across or $\alpha=\pi/3$ otherwise.

\begin{figure}[tb]
\centerline{\includegraphics[width=0.9\linewidth]{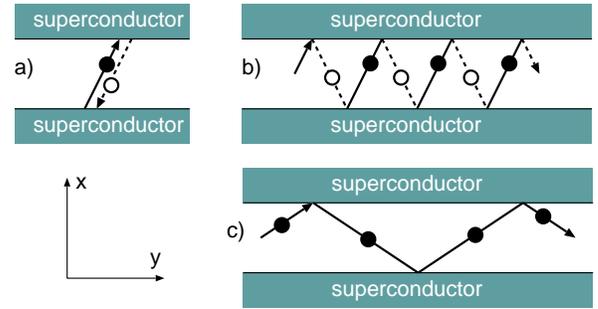}}
\caption{\label{Andreev_mode}
Three types of states in an SNS channel in graphene. The solid and
dashed lines show the classical trajectories of an electron (filled
circle) and a hole (open circle), converted into each other upon
Andreev reflection at the superconductor. The transition from a
localized Andreev level (a) to a propagating Andreev mode (b) occurs
when the excitation energy $\varepsilon$ becomes larger than the Fermi
energy $\mu$ in the normal region. These two types of states are both
charge-neutral. Purely electronic states (c) exist near grazing
incidence. While the states of type (a) and (b) are sensitive to the
phase difference $\phi$ of the two superconductors, the states of type
(c) are not.}
\end{figure}

The interface with a superconductor provides an altogether different
way to confine the carriers. At energies below the superconducting gap
$\Delta_{0}$, the electron and hole excitations in a
superconductor--normal-metal--superconductor (SNS) junction are
confined to the normal region. In usual metals this confinement leads
to bound states known as Andreev levels.\cite{And64,Kul70} They
consist of counterpropagating electrons and holes converted into each
other by Andreev retro-reflection at the NS boundaries (see Fig.\
\ref{Andreev_mode}a). Andreev levels carry an electrical current (a
supercurrent) across the NS interfaces, but they are
``quasi-localized'' along the interfaces. More precisely, the group
velocity of the Andreev levels along the NS interface is much smaller
than the Fermi velocity, and weak disorder fully localizes
them.\cite{Shy98}

As pointed out in Ref.\ \onlinecite{Bee06}, Andreev reflection in
undoped graphene is specular reflection instead of retro-reflection
(see Fig.\ \ref{Andreev_mode}b). The consequences were investigated in
that paper and in Ref.\ \onlinecite{Bha06} for a single NS
interface. Here we consider the consequences for an SNS junction.

We find that the transition from retro-reflection to specular
reflection is accompanied by a transition from quasi-localized Andreev
levels to propagating modes (``Andreev modes''), provided that
$E_{T}\ll\Delta_{0}$. This is the long-junction regime. (The states
remain localized in the opposite short-junction regime
$E_{T}\gg\Delta_{0}$, considered in Ref.\ \onlinecite{Tit06}.) The
transition (governed by the ratio $\mu/E_{T}$ of the Fermi energy
$\mu$ in N and the Thouless energy) has a drastic effect on the
density of states. While the excitation spectrum is gapless for
$\mu\gg E_{T}$, a gap opens up for $\mu\lesssim E_{T}$. The excitation
gap
\begin{equation}
E_{0}=\tfrac{1}{2}(\pi-|\phi|)E_{T}\label{E0SNS}
\end{equation}
has the same form as the band gap (\ref{E0armchair}) for confinement
by armchair edges --- with the phase difference $\phi\in(-\pi,\pi)$ of
the two superconductors taking over from the crystallographic phase
$\alpha$.

The Andreev modes have the same dispersion relation as the ``armchair
modes'' for confinement by armchair edges, and they are also
constructed out of states taken from two different valleys in the
Brillouin zone. However, while the armchair modes contain either
electron states from the conduction band or hole states from the
valence band, the Andreev modes are a superposition of conduction and
valence band states. As a consequence, the Andreev modes transport
energy but no charge along the NS interface --- so they will play a
role in thermal conduction along the interface but not in electrical
conduction.

The outline of this paper is as follows. The modes propagating along
the channel are characterized by their dispersion relation in Sec.\
\ref{dispersionDOS}. Both exact numerical and approximate (but highly
accurate) analytical results are given. From the dispersion relation
we determine the excitation gap in Sec.\ \ref{gap} and the density of states in Sec.\ \ref{DOS}, contrasting in
particular the low- and high-doping regimes. We derive the result
(\ref{E0SNS}) for the excitation gap in the low-doping regime and show
numerically that the gap closes $\propto E_0 (E_{T}/\mu)^{2}$ with increasing
doping. One way to measure the gap is by tunneling
spectroscopy. Another way, which we analyze in some detail in Sec.\
\ref{thermalG}, is by means of the thermal conductance of the channel
(for heat flow parallel to the NS boundaries). We conclude in Sec.\
\ref{conclude}.

\section{Dispersion relation}
\label{dispersionDOS}


\subsection{Quantization condition}
\label{quantcond}

To calculate the dispersion relation of the Andreev modes we solve the
Dirac-Bogoliubov-De Gennes (DBdG) equation\cite{Bee06} for the pair
potential
\begin{equation}
\Delta(\bm{r})=\left\{\begin{array}{cl}
\Delta_{0}\exp(i\phi/2)&{\rm if}\;\;x<-d/2,\\
0&{\rm if}\;\;-d/2<x<d/2,\\
\Delta_{0}\exp(-i\phi/2)&{\rm if}\;\;x>d/2.
\end{array}\right.\label{Deltaxdependence}
\end{equation}
We seek plane wave solutions $\Psi(x,y)=\psi(x)e^{iqy}$, with $q$ the
component of the wave vector parallel to the NS interfaces at $x=\pm
d/2$. The excitation energy $\varepsilon>0$ of the mode is measured
relative to the Fermi energy $\mu$ in the normal region
$|x|<d/2$. (The superconducting regions are assumed to be heavily
doped, with Fermi energy $\mu'\gg\mu$.)

The dispersion relation follows from the quantization condition
derived from the DBdG equation in Ref.\ \onlinecite{Tit06},
\begin{eqnarray}
\cos\phi&=&\left(\cos\theta_{+}\cos\theta_{-}+\frac{\sin\theta_{+}\sin\theta_{-}}
{\cos\alpha_{+}\cos\alpha_{-}}\right)\cos 2\beta\nonumber\\
&&\mbox{}+\left(\frac{\sin\theta_{+}\cos\theta_{-}}{\cos\alpha_{+}}-
\frac{\cos\theta_{+}\sin\theta_{-}}{\cos\alpha_{-}}\right)\sin 2\beta\nonumber\\
&&\mbox{}-\sin\theta_{+}\sin\theta_{-}\tan\alpha_{+}\tan\alpha_{-}.\label{qc}
\end{eqnarray}
The three angles $\alpha_{\pm},\theta_{\pm},\beta$ are functions of 
$\varepsilon$ and $q$,
\begin{eqnarray}
&&\alpha_{\pm}=\arcsin\left(\frac{{q} }{\mu\pm\varepsilon}\right),
\;\;\theta_{\pm}=\frac{\mu\pm\varepsilon}{E_{T}}\cos\alpha_{\pm},
\label{alphathetadef}\\
&&\beta=\arccos(\varepsilon/\Delta_{0}).
\label{betadef}
\end{eqnarray}
The quantization condition is invariant under $\mu\rightarrow -\mu$, 
so without loss of generality we may take $\mu>0$.

While Ref.\ \onlinecite{Tit06} dealt with the short-junction regime
$E_{T}\gg\Delta_{0}$, here we are concerned with the long-junction
regime $E_{T}\ll\Delta_{0}$. (Since the Thouless energy
$E_{T}\equiv\hbar v/d$, the latter criterion is equivalent to the
requirement that the separation $d$ of the NS interfaces is large
compared to the superconducting coherence length $\xi\equiv\hbar
v/\Delta_{0}$.) We furthermore restrict ourselves to low-lying
excitations, $\varepsilon\ll\Delta_{0}$. The relative magnitude of
$\varepsilon$ and $E_{T}$ is arbitrary. For ease of notation we will
use units such that $\hbar v\equiv 1$ in the intermediate
calculations, restoring the units in the final results.

For low-lying excitations $\ep\ll \Delta_0$ the quantization condition
(\ref{qc}) simplifies to
\be
\label{qc1}
\cos\phi+\cos\theta_+\cos\theta_-+r \sin\theta_+\sin\theta_-=0,
\e
where we have abbreviated
\begin{equation}
r=\frac{1+\sin\alpha_+\sin\alpha_-}{\cos\alpha_+\cos\alpha_-}.\label{rdef}
\end{equation}
The solutions to this equation can be represented in the form
$\ep=\ep^{\pm}_n(q)$, where $n=0,1,2,\dots$ is the mode index due to
the quantization of the motion in the $x$-direction and the
superscript $\pm$ accounts for the different $\phi$-dependence of the
modes.

\subsection{Exact solution}
\label{exact}

\begin{figure}[tb]
\centerline{\includegraphics[width=0.8\linewidth]{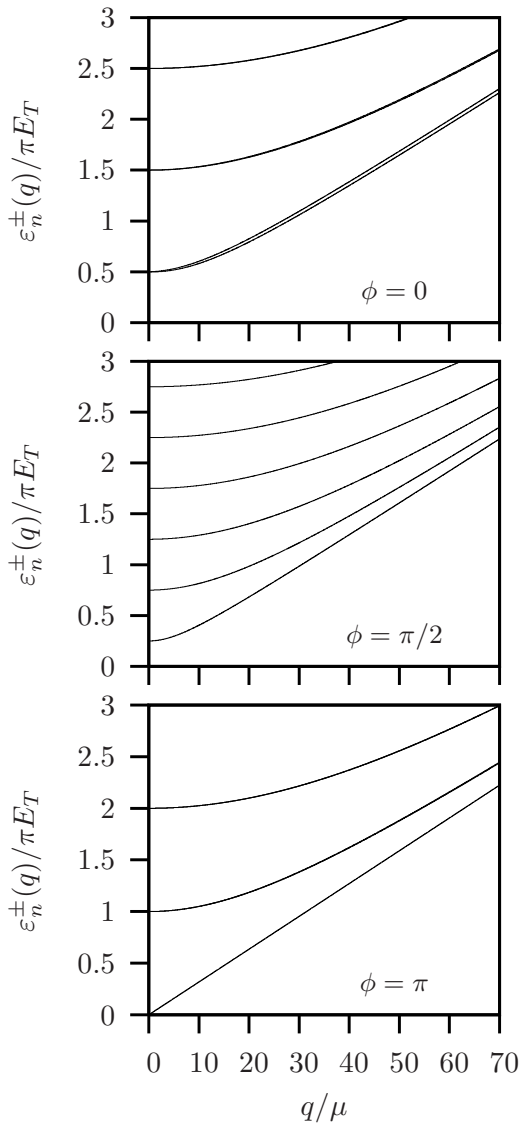}}\medskip
\caption{\label{figdispersionlowdoping}
Dispersion relation of the SNS junction, calculated numerically from
Eq.~(\ref{qc1}) for three values of the superconducting phase
difference $\phi$ at $\mu/E_{T}=0.1$. The lowest modes
$\ep_n^{\pm}(q)$ with $n=0,1,2$ are nearly degenerate for $\phi=0$
and nondegenerate for $\phi=\pi/2$ (thicker lines correspond to $\ep^+_n$).
For $\phi=\pi$ all modes are nearly degenerate except the lowest one
$\ep^-_0$.}
\end{figure}

\begin{figure}[tb]
\centerline{\includegraphics[width=0.8\linewidth]{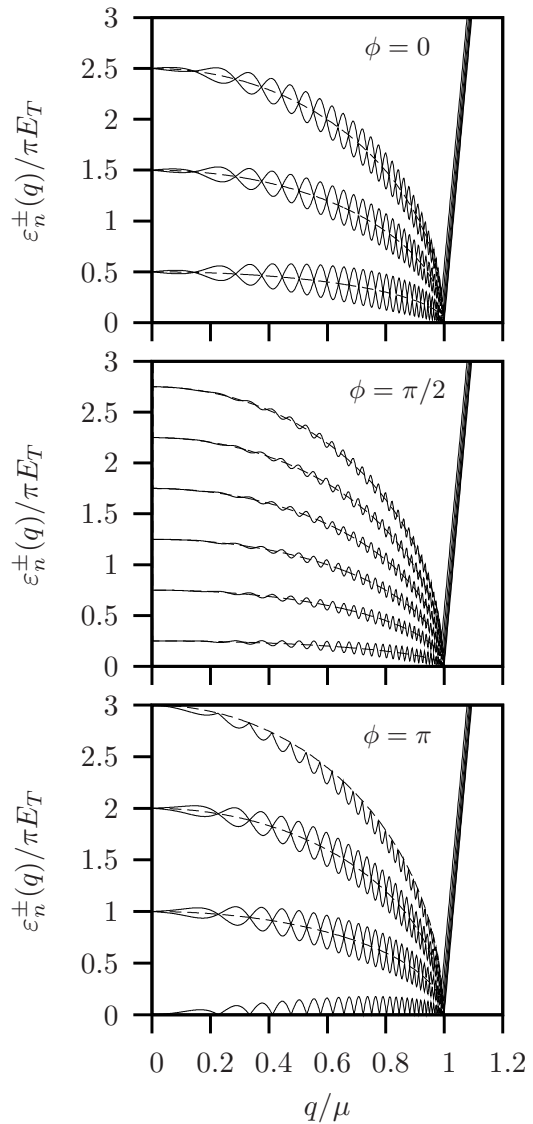}}\medskip
\caption{\label{figdispersionhighdoping}
Same as in Fig.~\ref{figdispersionlowdoping} for $\mu/E_T=100$.
The smoothed dispersion relation (\ref{barepsilon}) is indicated by dashed lines.}
\end{figure}

\begin{figure}[tb]
\centerline{\includegraphics[width=0.9\linewidth]{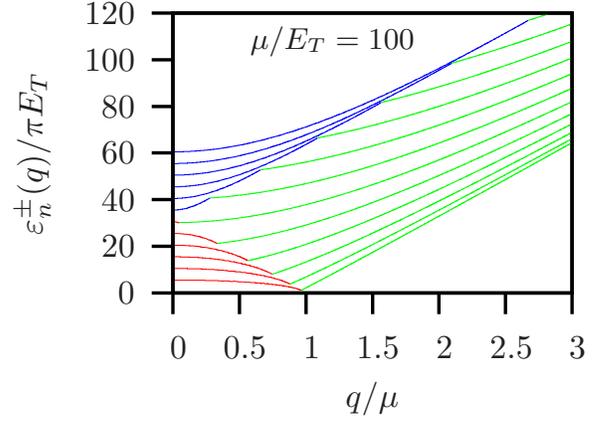}}\medskip
\caption{\label{figdispersiongeneral}
Dispersion relation of the SNS junction, calculated numerically from
Eq.\ (\ref{qc1}) for $\phi=0$ and $\mu/E_T=100$. The curves show
$\ep_n^{+}(q)$ with $n=5,10,15, \dots 60$. The three types of states
from Fig.\ \ref{Andreev_mode} are color coded; red=type a, blue=type
b, green=type c.} 
\end{figure}

The quantization condition (\ref{qc1}) can be solved
numerically. Results are shown in Figs.\ \ref{figdispersionlowdoping}, 
\ref{figdispersionhighdoping}, and \ref{figdispersiongeneral}. Only positive $q$ is plotted, because
$\varepsilon_{n}^{\pm}(-q)=\varepsilon_{n}^{\pm}(q)$.

The dispersion relation has three distinctly different branches,
indicated in Fig.\ \ref{figdispersiongeneral}, corresponding to the
three types of trajectories of Fig.\ \ref{Andreev_mode}.
\begin{itemize}
\item[(a)] The branch with $\hbar v|q|<\mu-\varepsilon$ (red)
describes {\em intra\/}band electron-hole states, corresponding to the
Andreev modes of Fig.\ \ref{Andreev_mode}a. The dispersion relation
for these modes has small oscillations as a function of $q$ on the
scale $1/d$, around a smooth convex curve (see Fig.\
\ref{figdispersionhighdoping}).  
\item[(b)] The branch with $\hbar
v|q|<\varepsilon-\mu$ (blue) describes {\em inter\/}band electron-hole
states, corresponding to the Andreev modes of Fig.\
\ref{Andreev_mode}b. The dispersion relation is concave without
oscillations.  
\item[(c)] The branch with $\hbar v|q|>|\ep-\mu|$
(green) corresponds to the purely electronic states of Fig.\
\ref{Andreev_mode}c. The hole component of the wave function can not
propagate along the channel because the reflection angle $\alpha_-$ of
the hole is imaginary on this branch. The dispersion relation on
branch (c) is concave, without oscillations, and joined to branch (a)
or (b) by a cusp singularity.
\end{itemize}

After these exact results we continue with an approximate, but highly
accurate, analytical solution of the quantization condition. We
consider separately the electron-hole modes with $\hbar
v|q|<|\varepsilon-\mu|$ and the electron modes with $\hbar
v|q|>|\varepsilon-\mu|$.

\subsection{Electron-hole modes}
\label{electronholemodes}

For $|q|<|\varepsilon-\mu|$ (setting again $\hbar v\equiv 1$) we
define the transverse momentum $p$ by the relation
\be
\theta_+-\theta_-=2p/E_{T}.
\e
The solution to this equation is given by
\be
\label{smallq}
\ep={p}\sqrt{1-\frac{{q}^2}{\mu^2-{p}^2}}.
\e
The condition $|q|<|\varepsilon-\mu|$ is equivalent to
$|q|<\lt|\mu-{p}^2/\mu\rt|\equiv q_{c}$. The momentum $q_{c}$ is the
location of the cusp in the dispersion relation, beyond which the hole
component of the mode vanishes.

If we express $\varepsilon$ in terms of $p$ with the help of Eq.\ (\ref{smallq}), we can write
\beml
\beq
&&r=\frac{(\mu^2-{p}^2)^2+(\mu {q})^2}{(\mu^2-{p}^2)^2-(\mu {q})^2},\\
&&\theta_++\theta_-=\frac{2\mu}{E_T}\sqrt{1-\frac{{q}^2}{\mu^2-{p}^2}}.
\eq
\eml
This allows to recast the quantization condition (\ref{qc1}) as
\beq
&&\label{qc2}\n
\cos\phi+\cos\lt(
\frac{2\mu}{E_T}\sqrt{1-\frac{{q}^2}{\mu^2-{p}^2}}\rt)\\
&&\;\;\;=\lt(\frac{\mu^2-{p}^2}{\mu {q}}\rt)^2\bigl[\cos\phi+\cos(2p/E_{T})\bigr],
\eq
which defines the quantization of the transverse momentum $p=p_{n}^{\pm}$. 

For $|q|\ll q_{c}$ the solution to Eq.\ (\ref{qc2}) is given by
\be
\label{averaged}
p_n^{{\pm}}=\pi E_{T}\lt(n+\frac{1}{2}\pm\frac{\phi}{2\pi}\rt),
\e
with $n=0,1,2,\dots$ and $\phi\in(-\pi,\pi)$. As $q$ approaches the
cusp at $q_{c}$ the first term in Eq.\ (\ref{qc2}) causes the
dispersion relation $\varepsilon(q)$ to oscillate rapidly around a
smooth curve $\bar{\varepsilon}(q)$. This smoothed dispersion relation
is obtained by substitution of Eq.\ (\ref{averaged}) into Eq.\
(\ref{smallq}), resulting in
\begin{eqnarray}
\bar{\varepsilon}^{\pm}_{n}&=&\pi E_{T}\lt(n+\frac{1}{2}\pm\frac{\phi}{2\pi}\rt)\nonumber\\
&&\mbox{}\times\sqrt{1-\frac{{q}^2}{\mu^2-(\pi E_{T})^{2}\lt(n+\frac{1}{2}\pm\phi/2\pi\rt)^2}},\nonumber\\
&&|q|<q_{c}=\left|\mu-\frac{(\pi E_{T})^{2}}{\mu}\lt(n+\frac{1}{2}\pm\frac{\phi}{2\pi}\rt)^2\right|.\label{barepsilon}
\end{eqnarray}
The smoothed dispersion for the lowest modes 
is indicated in Fig.~\ref{figdispersionhighdoping}
by a dashed line.

To determine also the rapid oscillations, we proceed as follows. The
quantization condition for $p$ can be written as
\be
\label{pgamma}
p_n^{\pm}=\pi E_{T}\lt(n+\frac{1}{2}\pm\frac{\gamma^{\pm}_n}{2\pi}\rt),
\e
where the phases $\gamma_{n}^\pm$ can be determined by iteration from
Eq.\ (\ref{qc2}). The first iteration turns out to be already highly
accurate in the high-doping regime $\mu \gg E_T$. It is given by
\beq
\n
&&\gamma^{\pm}_{n}=\arccos 
\lt[\lt(1-\frac{(\mu{q})^2 }{(\mu^2-{p}^2)^2}\rt)\cos\phi\rt.\\
&&
-\lt.\frac{(\mu{q})^2}{(\mu^2-{p}^2)^2}
\cos\lt(\frac{2\mu}{E_T}\sqrt{1-\frac{{q}^2}{\mu^2-{p}^2}}\rt)\rt],
\label{gamma}
\eq
where the momentum $p$ on the right-hand-side is taken in the zeroth
approximation (\ref{averaged}). The difference between the approximate
analytical results of Eqs.\ (\ref{smallq},\ref{pgamma},\ref{gamma})
and the exact numerical results plotted in Fig.\
\ref{figdispersionhighdoping} are not visible on the scale of that
figure.

\subsection{Electron modes}
\label{electronmodes}

For $|q|>|\varepsilon-\mu|$ the angle $\theta_-$ becomes strictly imaginary.
In this interval we define the transverse momentum $p$ by 
\be
\label{pdef2}
\theta_+=2 p/E_T.
\e
The condition $|q|>|\varepsilon-\mu|$ is then still equivalent to
$|q|>|\mu-{p}^2/\mu|\equiv q_{c}$. From Eq.~(\ref{pdef2}) we cast
the branch $|q|>q_c$ of the dispersion relation in the form
\be
\label{largeq}
\ep=\sqrt{{q}^2+4{p}^2}-\mu,
\e
where the momentum $p=p^{\pm}_n$ is quantized. The exact quantization 
condition follows directly from Eq.~(\ref{qc1}). 

For large longitudinal momenta $|q| \gg {\rm max}\lt(E_T, q_c\rt)$
the reflection angle $\theta_-\approx \pm i q/E_T$ of the hole takes on large
imaginary values. 
Therefore both $\sin\theta_-$ and $\cos\theta_-$ in
the quantization condition (\ref{qc1}) are exponentially large and the
$\phi$-dependence of the solution can be neglected.
This shows that the electron modes 
are insensitive to the superconducting phase difference across the channel.

At the cusp $|q|=q_c$ of the disperrsion relation we find $\theta_-=0$ and $\varepsilon=p^2/\mu$.
The coefficient $r$ in Eq.~(\ref{qc1}) tends to infinity, leading to 
\be
\lim_{|q|\rightarrow q_{c}}
r \sin\theta_-  = \frac{\mu^2-p^2}{p E_T}.
\e 
The quantization condition at the cusp 
thus simplifies to
\be
\label{qc3}
\cos\phi+\cos (2 p/E_T)=
\frac{\mu}{E_T} \left(\frac{p}{\mu}-\frac{\mu}{p}\right)\sin (2 p/E_T).
\e
For $\mu\gg E_T$ the condition is reduced to $\sin (2 p/E_T) =0$ with solution
\be
\label{ppi}
p^+_n=\pi E_{T}(n+1),\qquad p^-_n=\pi E_{T}\lt(n+\tfrac{1}{2}\rt),
\e
which is, again, $\phi$-independent. The quantization condition (\ref{ppi})
formally corresponds to $\gamma_n^{+}=\pi$, $\gamma_n^{-}=0$ in Eq.~(\ref{pgamma}).

\begin{figure}[tb]
\centerline{\includegraphics[width=0.9\linewidth]{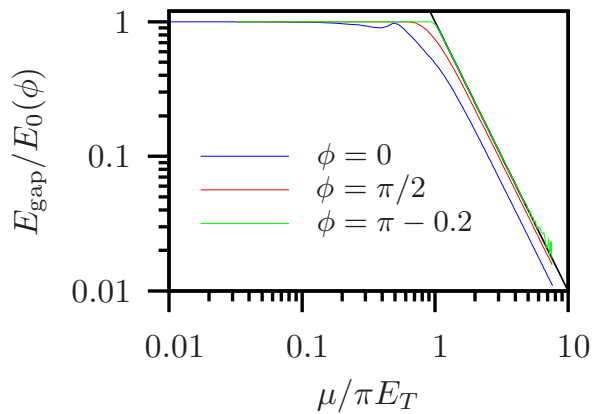}}\medskip
\caption{\label{figgap}
Double-logarithmic plot of the energy dependence of the excitation gap, calculated numerically 
from Eq.~(\ref{qc1}), for three different values of the superconducting phase difference
$\phi$ (colored lines). The straight black line is the asymptote $E_{\rm gap}/E_0(\phi)\propto(\pi E_T/\mu)^2$.
}
\end{figure}

\section{Excitation gap}
\label{gap}

At small doping $\mu \lesssim E_T$ the excitation gap is given by
\begin{equation}
E_{\rm gap}=\tfrac{1}{2}(\pi-|\phi|)E_T\equiv E_0(\phi),
\end{equation} 
which is the energy 
of the lowest mode at $q=0$. For $\mu \gtrsim E_T$ the gap 
is given by the energy of the lowest mode at a nonzero longitudinal 
momentum $|q|\lesssim q_c$, which corresponds to the deepest minimum of the oscillatory 
dispersion relation. We have not succeeded in determining this minimum analytically from the quantization condition (\ref{qc1}), but we have a very accurate numerical solution. 

Results for different values of $\phi$ are presented in Fig.~\ref{figgap}. 
One can see that the ratio $E_{\rm gap}/E_0(\phi)$
depends only weakly on the superconducting phase difference $\phi$
and that the crossover to a decay $\propto\mu^{-2}$ happens in a narrow interval 
around $\mu = \pi E_T$. As shown by the black 
line in Fig.~\ref{figgap}, the large-$\mu$ asymptote is given by
\begin{equation} 
E_{\rm gap}=c(\phi)E_0(\phi)(\pi E_T/\mu)^2,
\end{equation} 
with $c(\phi)$ increasing from $1/2$ at $\phi=0$ to $1$ at $|\phi|=\pi$. 

\begin{figure}[tb]
\centerline{\includegraphics[width=0.9\linewidth]{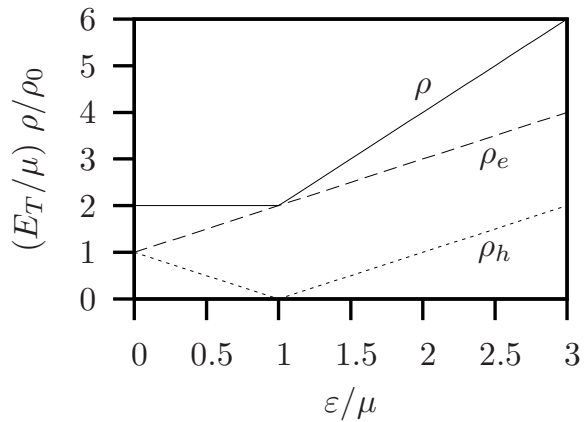}}
\caption{\label{fig_RHOTHERMO}
Thermodynamic limit $E_{T}=\hbar v/d\rightarrow 0$ of the density 
of states $\rho$ of the SNS junction, according to Eq.\ (\ref{RHOTHERMO}) 
with $\rho_{0}=4{\cal L}/\pi\hbar v$. The total density of states 
$\rho$ (solid) is the sum of the density $\rho_{e}\propto\mu+\varepsilon$ 
of electron states (dashed) and the density $\rho_{h}\propto|\mu-\varepsilon|$ 
of hole states (dotted).
}
\end{figure}

\section{Density of states}
\label{DOS}

\subsection{Thermodynamic limit}
\label{thermolimit}

Before turning to the calculation of the density of states at finite $E_{T}$, 
it is instructive to first determine the behavior in the thermodynamic 
limit $d\rightarrow\infty\Leftrightarrow E_{T}\rightarrow 0$. The DBdG 
equation then decouples into separate Dirac equations for electrons and holes. 
The total density of states 
$\rho(\varepsilon)=\rho_{e}(\varepsilon)+\rho_{h}(\varepsilon)$ 
is the sum of the electron density of states $\rho_{e}(\varepsilon)\propto\mu+\varepsilon$ 
and the hole density of states $\rho_{h}(\varepsilon)\propto|\mu-\varepsilon|$, leading to
\begin{equation}
\rho(\varepsilon)=\frac{4{\cal L}d}{\pi(\hbar v)^{2}}\max(\mu,\varepsilon),
\;\;{\rm if}\;\;\varepsilon\gg E_{T}.\label{RHOTHERMO}
\end{equation}
Here ${\cal L}$ is the extension of the junction in the $y$-direction
and the factor of $4$ accounts for the spin and valley
degeneracies. (For a derivation of Eq.\ (\ref{RHOTHERMO}) directly
from the quantization condition (\ref{qc1}), see App.\
\ref{derivationthermo}.)

In Fig.\ \ref{fig_RHOTHERMO} we plot the density of states $\rho$ of
the DBdG equation together with the separate electron and hole
contributions $\rho_{e}$ and $\rho_{h}$. The superconducting proximity
effect will introduce fine structure in $\rho$ on the scale of the
Thouless energy $E_{T}$, as we will determine in the next
subsections. We consider separately the low-doping regime $\mu\ll
E_{T}$, where the contribution from {\em inter\/}band electron-hole
modes dominates, and the high-doping regime $\mu\gg E_{T}$, where the
{\em intra\/}band electron-hole modes dominate.

\subsection{Low-doping regime}
\label{lowdopingregime}

To determine the excitation spectrum in the low-doping regime, we take
the $\mu\rightarrow 0$ limit of Eq.\ (\ref{barepsilon}), resulting in
\begin{equation}
\varepsilon^{\pm}_{n}=\sqrt{(\hbar vq)^{2}+(\pi E_{T})^{2}
(n+\tfrac{1}{2}\pm\phi/2\pi)^2},\label{dispersionlowdoping}
\end{equation}
for $n=0,1,2,\ldots$ and $\phi\in(-\pi,\pi)$. (There is no need to
distinguish $\bar{\varepsilon}$ from $\varepsilon$, because the
dispersion relation does not oscillate in this regime.) The two series
of modes $\varepsilon_{n}^{+}$ and $\varepsilon_{n}^{-}$ are
nondegenerate, except for $\phi=0,\pi$. (The lowest mode
$\varepsilon_{0}^{-}$ is nondegenerate also for $\phi=\pi$.)

\begin{figure}[tb]
\centerline{\includegraphics[width=0.9\linewidth]{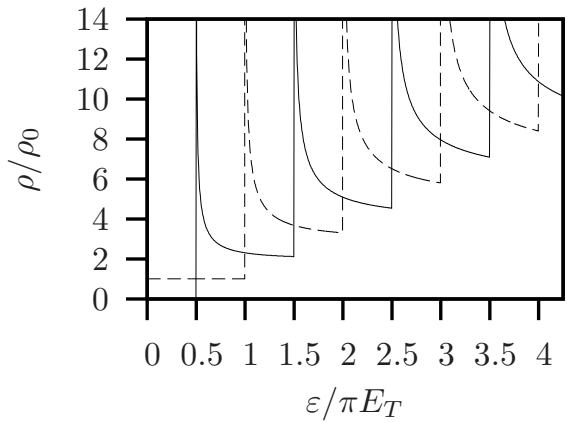}}
\caption{\label{figlowdoping}
Density of states of the SNS junction in the low-doping regime, for
superconducting phase difference $\phi=0$ (solid curves) and
$\phi=\pi$ (dashed curves). The curves are calculated from Eq.\
(\ref{rholowdoping}), normalized by $\rho_{0}=4{\cal L}/\pi\hbar
v$. The excitation gap for $\phi=0$ is at $E_{0}=\pi E_{T}/2$.  }
\end{figure}

In Fig.\ \ref{figlowdoping} we plot the density of states
\begin{eqnarray}
\rho(\varepsilon)&=&\frac{4\cal L}{\pi}\sum_{n=0}^{\infty}
\sum_{\pm}\left|\frac{\partial\varepsilon^{\pm}_{n}}{\partial q}\right|^{-1}\nonumber\\
&=&\frac{4{\cal L}\varepsilon}{\pi\hbar vE_{T}}\sum_{n=0}^{\infty}
\sum_{\pm}(X^{\pm}_{n})^{-1/2}\Theta(X^{\pm}_{n}),\label{rholowdoping}\\
X^{\pm}_{n}&=&(\varepsilon/E_{T})^{2}-\pi^{2}(n+\tfrac{1}{2}\pm\phi/2\pi)^2,
\label{Xdef}
\end{eqnarray}
with $\Theta$ the unit step function. The excitation spectrum has a 
gap at the energy $E_{0}$ given by Eq.\ (\ref{E0SNS}). The gap closes 
for $|\phi|=\pi$, when $\rho=4{\cal L}/\pi\hbar v\equiv\rho_{0}$ is 
constant at low energies. At large excitation energies $\varepsilon\gg E_{T}$ 
the sum over $n$ in Eq.\ (\ref{rholowdoping}) may be replaced by an integral, 
resulting in a linearly increasing density of states,
\begin{equation}
\rho(\varepsilon)=\frac{4{\cal L}\varepsilon}{\pi\hbar vE_{T}},\;\;{\rm if}
\;\;\varepsilon\gg E_{T},\label{rholargeepsilonsmallmu}
\end{equation}
in agreement with the thermodynamic limit (\ref{RHOTHERMO}).

The group velocity $v_{n}^{\pm}$ in the $y$-direction of 
the $n$-th mode is given by the derivative
\begin{equation}
v_{n}^{\pm}=\frac{\partial\varepsilon_{n}^{\pm}}{\hbar\partial q}.
\label{vlowdoping}
\end{equation}
For each propagating mode $v_{n}^{\pm}\rightarrow v$ with increasing
excitation energy. These are all interband electron-hole modes. The
purely electronic modes are pushed to $|q|\gtrsim
E_{T}^{2}/\mu\rightarrow\infty$ in the low-doping regime
$\mu/E_{T}\rightarrow 0$, while the intraband electron-hole modes can
not propagate if $\varepsilon>\mu$.

\begin{figure}[tb]
\centerline{\includegraphics[width=0.9\linewidth]{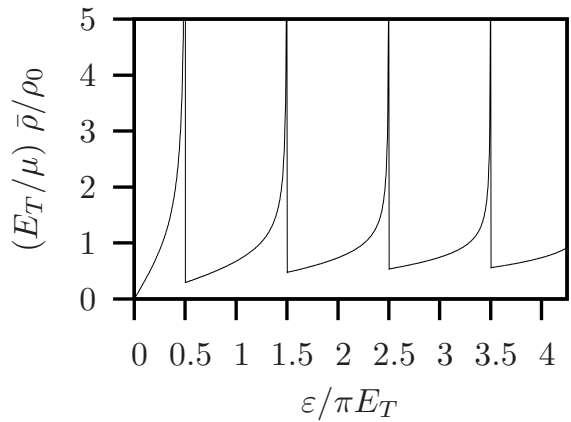}}
\caption{\label{figrhohighdoping}
Smoothed density of states of the SNS junction in the high-doping regime 
for $\phi=0$, calculated from Eq.\ (\ref{rhobar}).
}
\end{figure}

\subsection{High-doping regime}
\label{highdopingregime}

As shown in Sec.\ \ref{electronholemodes}, the electron-hole branch of
the dispersion relation for $\mu\gg E_{T}$ is a rapidly oscillating
function. Small local variations in the separation $d$ of the NS
interfaces, on the scale of the Fermi wave length
$\lambda_{F}=hv/\mu$, will average out these oscillations, leaving the
smoothed dispersion relation (\ref{barepsilon}). In the large-$\mu$
limit this reduces to
\begin{equation} \bar{\varepsilon}_{n}^{\pm}=\frac{\pi
E_{T}}{\mu}\left(n+\frac{1}{2}\pm\frac{\phi}{2\pi}\right)\sqrt{\mu^{2}-(\hbar
vq)^{2}}. 
\label{dispersionhighdopingsmooth} 
\end{equation}
The branch of purely electronic states (for $\hbar v|q|>\mu$) is not
described by Eq.\ (\ref{dispersionhighdopingsmooth}), but since it
contributes negligibly to the density of states for
$\varepsilon\ll\mu$ we need not consider it here.

The smoothed density of states is given by
\begin{eqnarray}
\bar{\rho}(\varepsilon)&=&\frac{4\cal L}{\pi}
\sum_{n=0}^{\infty}\sum_{\pm}\left|\frac{\partial\bar{\varepsilon}^{\pm}_{n}}{\partial q}\right|^{-1}\nonumber\\
&=&\frac{4{\cal L}\mu\varepsilon}{\pi^{2}\hbar vE^{2}_{T}}
\sum_{n=0}^{\infty}\sum_{\pm}(Y_{n}^{\pm})^{-1/2}\Theta(Y_{n}^{\pm})\nonumber\\
&&\qquad\qquad\mbox{}\times(n+\tfrac{1}{2}\pm\phi/2\pi)^{-1},\label{rhobar}\\
Y_{n}^{\pm}&=&\pi^{2}(n+\tfrac{1}{2}\pm\phi/2\pi)^{2}-(\varepsilon/E_{T})^{2}.
\label{Ydef}
\end{eqnarray}
We plot it in Fig.\ \ref{figrhohighdoping} for $\phi=0$.

The peaks in the density of states at $\varepsilon_{n}^{\pm}=\pi
E_{T}(n+\tfrac{1}{2}\pm\phi/2\pi)$ are analogous to the De
Gennes-Saint James resonances\cite{DeG63} in conventional Josephson
junctions. The lowest resonance is at the same energy
$E_{0}=\tfrac{1}{2}(\pi-|\phi|)E_{T}$ as the gap (\ref{E0SNS}) in the
low-doping regime --- however, in the high-doping regime the density
of states is gapless, vanishing linearly at small excitation energies
with a $\phi$-dependent slope:
\begin{equation}
\bar{\rho}(\varepsilon)=\frac{4{\cal L}\mu\varepsilon}{\pi\hbar vE_{T}^{2}}\,\frac{1}{\cos^{2}(\phi/2)},\;\;
{\rm if}\,\,\varepsilon\ll E_{0}.\label{rhosmallepsilon}
\end{equation}
The slope diverges when $\phi\rightarrow\pi$, because then the lowest
resonance is at $\varepsilon=0$. At high excitation energies
$\varepsilon\gg E_{T}$ (but still $\varepsilon\ll\mu$) the density of
states approaches a $\phi$-independent limit,
\begin{equation}
\bar{\rho}(\varepsilon)=\frac{4{\cal L}\mu}{\pi\hbar v E_{T}},\;\;{\rm if}\;\;
\varepsilon\gg E_{T},\label{rholargeepsilon}
\end{equation}
in agreement with Eq.\ (\ref{RHOTHERMO}).

\begin{figure}[tb]
\centerline{\includegraphics[width=0.8\linewidth]{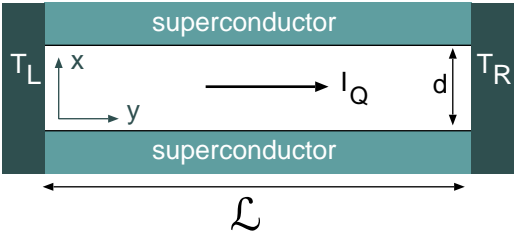}}
\caption{\label{figthermosetup}
A temperature difference $\Delta T=T_{L}-T_{R}$ between the two ends 
of the graphene channel drives a heat current $I_{Q}$, carried by 
Andreev modes in the normal region at temperatures below the gap $\Delta_{0}$ 
in the superconductors.
}
\end{figure}

The group velocity
$\bar{v}^{\pm}_{n}=\partial\bar{\varepsilon}^{\pm}_{n}/\hbar\partial
q$ corresponding to the smoothed density of states is of order
$vE_{T}/\mu\simeq v\lambda_{F}/d$ for $\hbar v|q|\ll\mu$, much smaller
than the carrier velocity $v$. This as expected from the classical
trajectories in Fig.\ \ref{Andreev_mode}a.

\section{Thermal conductance}
\label{thermalG}

The thermal conductance $G_{\rm thermal}=I_{Q}/\Delta T$ of the graphene channel, 
for heat flow $I_{Q}$ parallel to the NS boundaries, can be measured by applying a 
temperature difference $\Delta T=T_{L}-T_{R}$ between the two ends of the channel 
(see Fig.\ \ref{figthermosetup}). Experiments of this type have been performed in 
metals by Eom, Chien, and Chandrasekhar\cite{Eom98} and analyzed theoretically
in Refs.~\onlinecite{Bez03,Vir04}.

To determine the thermal conductance of the graphene channel 
we start from the Landauer-type formula\cite{But90,Hou92}
\begin{equation}
G_{\rm thermal}=-\frac{4}{2\pi\hbar T_{0}}\int_{0}^{\infty}d\varepsilon\,
\varepsilon^{2}\frac{\partial f}{\partial\varepsilon}\sum_{n}{\cal T}_{n}(\varepsilon),
\label{Gthermaldef}
\end{equation}
valid for small temperature differences $\Delta T\ll
T_{L},T_{R}$. (The factor of $4$ is again from the spin and valley
degeneracy.) We assume that the mean temperature
$T_{0}=(T_{L}+T_{R})/2$ is much less than $\Delta_{0}/k_{B}$, so that
the thermal current through the superconductors is suppressed
exponentially.\cite{And64} The function
$f(\varepsilon)=[1+\exp(\varepsilon/k_{B}T_{0})]^{-1}$ is the Fermi
function and ${\cal T}_{n}$ is the transmission probability of the
$n$-th propagating mode along the channel. In a ballistic channel each
of the $N(\varepsilon)$ propagating modes at energy $\varepsilon$ has
transmission probability ${\cal T}_{n}=1$, so we obtain the thermal
conductance
\begin{equation}
G_{\rm thermal}=\frac{1}{2\pi\hbar k_{B}T^{2}_{0}}\int_{0}^{\infty}d\varepsilon\,
\frac{\varepsilon^{2}N(\varepsilon)}{\cosh^{2}(\varepsilon/2k_{B}T_{0})}.
\label{Gthermal}
\end{equation}

\begin{figure}[tb]
\centerline{\includegraphics[width=0.8\linewidth]{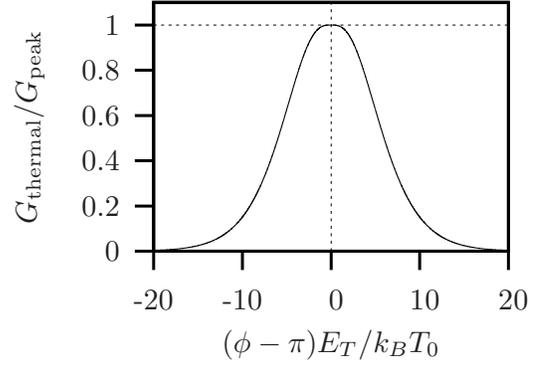}}
\caption{\label{figGthermallowdoping}
Thermal conductance of the SNS junction in the low-doping, low-temperature regime 
($\mu,k_{B}T_{0}\ll E_{T}$), calculated as a function of the superconducting phase 
difference $\phi$ (modulo $2\pi$) from Eq.\ (\ref{GthermallowdopinglowT}). 
The peak value equals $G_{\rm peak}=\pi k_{B}^{2}T_{0}/3\hbar$.
}
\end{figure}

In the low-doping, low-temperature limit $\mu,k_{B}T_{0}\ll E_{T}$
only the lowest mode contributes and the thermal conductance reduces
to
\begin{equation}
G_{\rm thermal}=\frac{k_{B}^{2}T_{0}}{2\pi\hbar}
\int_{E_{0}/k_{B}T_{0}}^{\infty}dx\,\frac{x^{2}}{\cosh^{2}(x/2)},
\label{GthermallowdopinglowT}
\end{equation}
with $E_{0}$ the gap given by Eq.\ (\ref{E0SNS}). As illustrated 
in Fig.\ \ref{figGthermallowdoping}, the thermal conductance in this low-doping, 
low-temperature regime vanishes unless the superconducting phase difference $\phi$ 
is in a narrow interval of order $k_{B}T_{0}/E_{T}$ around $\pi$ (modulo $2\pi$). 
The peak at $\phi=\pi$ has height
\begin{equation}
G_{\rm peak}=\frac{\pi}{3}\frac{k_{B}^{2}T_{0}}{\hbar}.\label{Gpeak}
\end{equation}

In the high-doping limit $\mu\gg E_{T}$ we may distinguish a
moderately-low temperature regime $E_{T}^{2}/\mu\ll k_{B}T_{0}\lesssim
E_{T}$ and an ultralow temperature regime $k_{B}T_{0}\ll
E_{T}^{2}/\mu$. In the ultralow temperature regime it is again only
the lowest mode which contributes, so Eq.\
(\ref{GthermallowdopinglowT}) remains valid if we replace $E_{0}$ by
$E_{\rm gap}$ from Sec.\ \ref{gap}. In the moderately-low temperature
regime there remains a large number
\begin{equation}
N(\varepsilon)= \frac{4}{\pi E_{T}}\sqrt{\varepsilon\mu}+{\cal O}(1)\label{Nlargemu}
\end{equation}
of modes that contributes at energies $\varepsilon\lesssim
k_{B}T_{0}$. Substitution into Eq.\ (\ref{Gthermal}) gives the thermal
conductance
\begin{equation}
G_{\rm thermal}=2.34\,\frac{k_{B}^{2}T_{0}}{\hbar}\sqrt{\frac{\mu k_{B}T_{0}}{E_{T}^{2}}}.\label{Gthermalhighdoping}
\end{equation}
The thermal conductance is insensitive to the superconducting phase
difference because of the vanishing excitation gap in the high-doping
regime.

\section{Conclusion}
\label{conclude}

We have shown that a graphene channel with superconducting boundaries
supports a type of propagating modes along the channel that do not
exist in usual SNS junctions. These ``Andreev modes'' exist because
the Andreev reflection close to the Dirac point of vanishing Fermi
energy $\mu$ is specular.\cite{Bee06} The Andreev modes are charge
neutral, so they transport energy but no charge along the channel.

The thermal conductance due to the Andreev modes depends strongly on
the superconducting phase difference $\phi$, because of the
$\phi$-dependent excitation gap $E_{0}$ of the Andreev modes. Away
from the Dirac point the character of the Andreev reflection changes
from specular reflection to retroreflection. The excitation gap closes
and the thermal conductance becomes $\phi$-independent.

The closing of the excitation gap with increasing doping can be
studied directly by point contact spectroscopy (tunneling into the
graphene layer via a tunnel probe on top of the layer).

\acknowledgments
This research was supported by the Dutch Science Foundation NWO/FOM
and by the German Science Foundation DFG through SFB 513. M.T. acknowledges
the hospitality in the Max-Planck-Institute for Physics of Complex Systems 
in Dresden. 

\appendix

\section{Derivation of Eq.\ (\protect\ref{RHOTHERMO}) from the quantization condition}
\label{derivationthermo}

It is instructive to calculate the density of states in the
thermodynamic limit directly from the quantization condition
(\ref{qc1}). The dispersion relation for all $q$ can be compactly
written as
\be
\label{dispersion_gen}
\ep^{\pm}_n(q)=\lt\{
\ba{ll} 
{\displaystyle{p}\sqrt{1-\frac{{q}^2}{\mu^2-{p}^2}},}&\quad {q}\leq |\mu-{p}^2/\mu|,\\
{\displaystyle\sqrt{4{p}^2+{q}^2}-\mu,}&\quad {q}\geq |\mu-{p}^2/\mu|,
\ea
\rt.
\e
where the momentum $p=p^{\pm}_n$ is quantized according to Eqs.\
(\ref{pgamma}) and (\ref{ppi}).  The density of states is given by
\be
\label{dos1}
\rho(\ep)=\frac{4 {\cal L}}{\pi}\s_{n=0}^\infty\s_{\pm}\int_0^\infty
dq\,\delta(\ep-\ep^{\pm}_n(q)).
\e

Since the quantization condition of the momentum $p$ is linear in the
mode index $n$, we can replace the summation over $n$ with the
integration over $p$ in the thermodynamic limit
$d\rightarrow\infty$. In this limit we can ignore the dependence of
the phases $\gamma^{\pm}_n$ on $q$. The integral Eq.\ (\ref{dos1})
results in
\beq
\n
\rho(\ep)&=&\frac{8{\cal L}d}{(\pi \hbar v)^2}
\il_{{\rm min}\lt\{\ep,\sqrt{\ep\mu}\rt\}}^{{\rm max}\lt\{\ep,\sqrt{\ep\mu}\rt\}}
d{p}\,
\frac{\ep}{{p}}\sqrt{\frac{\mu^2-{p}^2}{{p}^2-\ep^2}}\\
&+&
\frac{8{\cal L}d}{(\pi \hbar v)^2}\il_0^{\sqrt{\ep\mu}} 
d{p}\,
\frac{\ep+\mu}{\sqrt{(\ep+\mu)^2-4{p}^2}} \label{dos2}\\
&=&\frac{4{\cal L}d}{\pi(\hbar v)^2}\,{\rm max}\,(\ep,\mu).
\eq

The first and second integral on the right-hand-side of Eq.\
(\ref{dos2}) are, respectively, the contributions from electron-hole
and electron modes to the density of states in the thermodynamic
limit. Even though each integral is a non-trivial function of energy,
their sum reduces to the elementary result (\ref{RHOTHERMO}),
confirming the consistency of our analysis.


\begin{thebibliography}{99}
\bibitem{Sai98} R. Saito, G. Dresselhaus, and M. S. Dresselhaus, 
{\em Physical Properties of Carbon Nanotubes\/} (Imperial College, London, 1998).
\bibitem{Per06} N. M. R. Peres, A. H. Castro Neto, and F. Guinea,
Phys.\ Rev.\ B {\bf 73}, 195411 (2006).
\bibitem{Bre06} L. Brey and H. A. Fertig, Phys.\ Rev.\ B {\bf 73}, 235411 (2006).
\bibitem{And64} A. F. Andreev, Sov.\ Phys.\ JETP {\bf 19}, 1228 (1964)
\bibitem{Kul70} I. O. Kulik, Sov.\ Phys.\ JETP {\bf 30}, 944 (1970).
\bibitem{Shy98} A. V. Shytov, P. A. Lee, and L. S. Levitov, Phys.\ Uspekhi {\bf 41}, 207 (1998).
\bibitem{Bee06} C. W. J. Beenakker, Phys.\ Rev.\ Lett.\ {\bf 97}, 067007 (2006).
\bibitem{Bha06} S. Bhattacharjee and K. Sengupta, cond-mat/0607489.
\bibitem{Tit06} M. Titov and C. W. J. Beenakker, Phys.\ Rev.\ B {\bf 74}, 041401(R) (2006).
\bibitem{DeG63} P. G. de Gennes and D. Saint-James, Phys.\ Lett.\ {\bf 4}, 151 (1963).
\bibitem{Eom98} J. Eom, C.-J. Chien, and V. Chandrasekhar, Phys.\ Rev.\ Lett.\ {\bf 81}, 437 (1998).
\bibitem{Bez03} E. V. Bezuglyi and V. Vinokur, Phys.\ Rev.\ Lett.\ {\bf 91}, 137002 (2003).
\bibitem{Vir04} P. Virtanen and T. T. Heikkil\"a, Phys.\ Rev.\ Lett.\ {\bf 92}, 177004 (2004).
\bibitem{But90} P. N. Butcher, J. Phys.\ Condens.\ Matt.\ {\bf 2}, 4869 (1990).
\bibitem{Hou92} H. van Houten, L. W. Molenkamp, C. W. J. Beenakker, and C. T. Foxon, 
Semicond.\ Science Techn.\ {\bf 7}, B215 (1992).

\end{thebibliography}
\end{document}